\documentclass{article}
\usepackage{arxiv}

\usepackage[utf8]{inputenc} % allow utf-8 input
\usepackage[T1]{fontenc}    % use 8-bit T1 fonts
\usepackage{hyperref}       % hyperlinks
\usepackage{url}            % simple URL typesetting
\usepackage{booktabs}       % professional-quality tables
\usepackage{amsfonts}       % blackboard math symbols
\usepackage{nicefrac}       % compact symbols for 1/2, etc.
\usepackage{microtype}      % microtypography
\usepackage{lipsum}		% Can be removed after putting your text content
\usepackage{graphicx}
\usepackage{doi}

\usepackage{algorithm}
\usepackage{algpseudocode}
\usepackage{subfigure}
\usepackage{stackengine}
\usepackage{amsmath}

\title{Scatter Ptychography}

\author{ 
    {Qian Huang}\thanks{Qian Huang is a student at Department of Electrical and Computer Engineering, Duke University, Durham, NC 27708. This work was finished when he was doing an internship at the University of Arizona.} \\
	Wyant College of Optical Sciences\\
	University of Arizona\\
	Tucson, AZ 85721 \\
	\texttt{qh38@email.arizona.edu} \\
	\And
    {Zhipeng Dong} \\
	Wyant College of Optical Sciences\\
	University of Arizona\\
	Tucson, AZ 85721 \\
	\texttt{zhipengdong@email.arizona.edu} \\
	\And
    \href{https://orcid.org/0000-0003-3011-9320}{\includegraphics[scale=0.06]{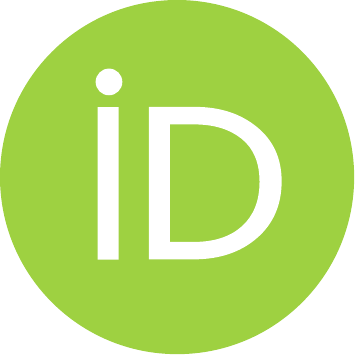}\hspace{1mm} Yuzuru Takashima} \\
	Wyant College of Optical Sciences\\
	University of Arizona\\
	Tucson, AZ 85721 \\
	\texttt{ytakashima@optics.arizona.edu} \\
	\And
    {Timothy J. Schulz} \\
	Department of Electrical and Computer Engineering\\
	Michigan Technological University\\
	Houghton, MI 49931 \\
	\texttt{schulz@mtu.edu} \\
	\And
    \href{https://orcid.org/0000-0001-5655-2478}{\includegraphics[scale=0.06]{orcid.pdf}\hspace{1mm} David J. Brady} \\
	Wyant College of Optical Sciences\\
	University of Arizona\\
	Tucson, AZ 85721 \\
	\texttt{djbrady@arizona.edu}
}

\begin{document}
\maketitle

\begin{abstract}
 Coherent illumination reflected by a remote target may be secondarily scattered by intermediate objects or materials. Here we show that phase retrieval on remotely observed images of such scattered fields enables imaging of the illuminated object at resolution proportional to $\lambda R_s/A_s$, where $R_s$ is the range between the scatterer and the target and $A_s$ is the diameter of the observed scatter. This resolution may exceed the resolution of directly viewing the target by the factor $R_cA_s/R_sA_c$, where $R_c$ is the range between the observer and the target and $A_c$ is the observing aperture. Here we use this technique to demonstrate $\approx 32\times$ resolution improvement relative to direct imaging\footnote{Measurement data, simulation code and reconstruction code presented in this manuscript is available online~\cite{code}.}. 
\end{abstract}

% keywords can be removed
% \keywords{First keyword \and Second keyword \and More}

\section{Background}

Phase retrieval consists of estimation of complex-valued fields from irradiance measurements\cite{fienup2013phase, fienup1982phase, Gerchberg72}. Typically, phase retrieval is applied where more direct observation is impossible. However, as proposed and demonstrated here, image transfer through scattering and phase retrieval may be used to substantially increase effective aperture relative to direct-view focal or holographic imaging. The proposed technique consists of imaging the irradiance of coherent illumination scattered by remote targets. The irradiance is insensitive to phase distortions imparted by the scattering process but phase retrieval on this signal enables imaging of the target by computational backprojection. 

Phase estimation is improved by observing the irradiance of the scattered field over multiple relative displacements of the target and scatter positions. In using such displacements to recover the target field the proposed approach builds on ptychographic methods based on lateral displacements of object fields or their Fourier transform~\cite{rodenburg2008ptychography,zheng2013wide,konda2020fourier, zheng2021concept}. With this in mind, we introduce the term "scatter ptychography" to describe the proposed method. 

While only laboratory experiments are presented here, we anticipate that scatter ptychography will be applicable to remote sensing through scattering and turbulence. Most previous efforts in such applications have focused on overcoming effective aperture distortion using adaptive optics or lucky imaging~\cite{law2009getting}. Some studies, however, considered the use of anisoplanatic effects to use the atmosphere to increase effective aperture size~\cite{Vorontsov:01}. The goal of scatter ptychography is also to use interactions with media along the optical path to increase aperture, but in contrast with coherent lensing effects, scatter ptychography relies on incoherently scattered signals. Scatter has previously been used to achieve super-resolution in radar imaging~\cite{chen1998experimental} but to our knowledge this has not previously been achieved at optical frequencies. Along with challenges associated with the larger data load of optical imaging, the primary difference is that phase retrieval must be used to recover the optical field. 

\begin{figure}[htbp]
    \centering
    \includegraphics[width=.7\linewidth]{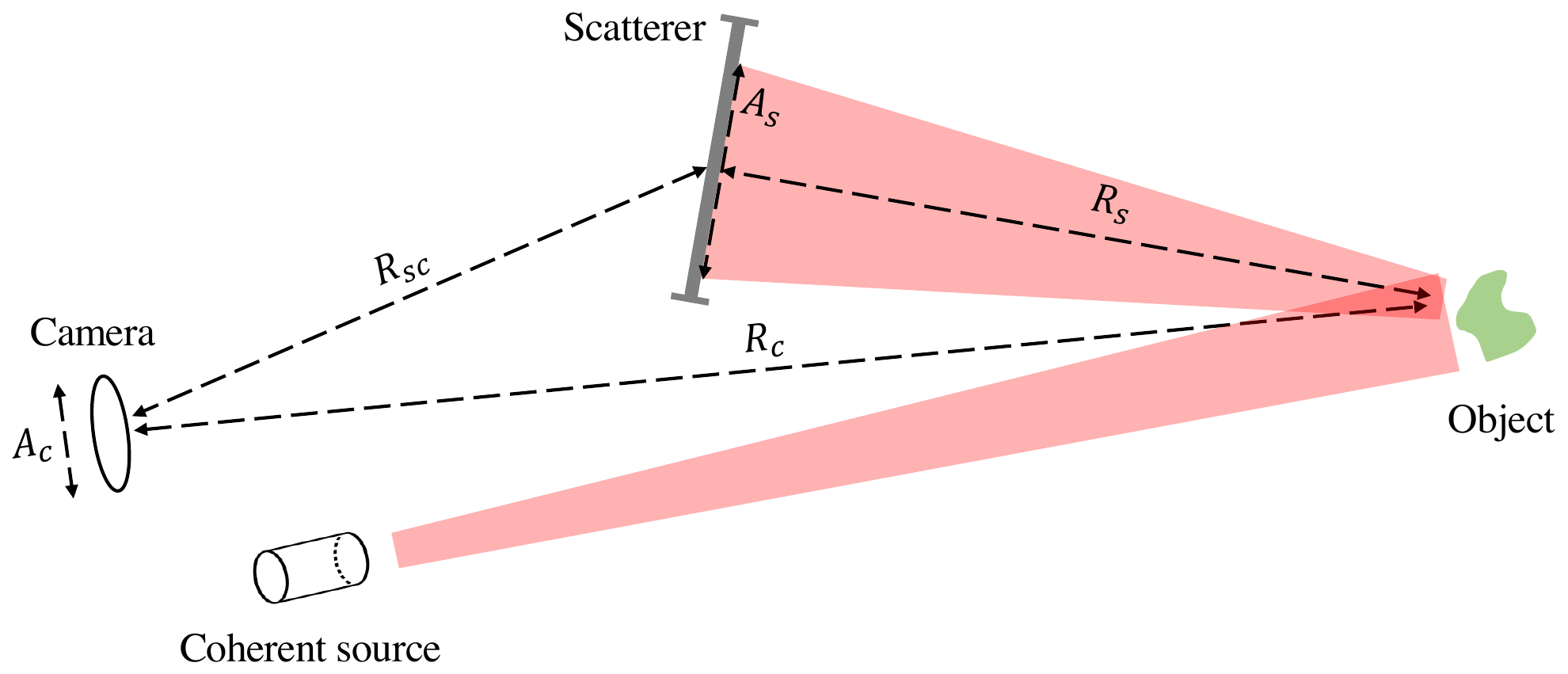}
    \caption{Scatter ptychography geometry.}
    \label{fig:systemGeometry}
\end{figure}

The basic geometry of scatter ptychography is shown in Fig.~\ref{fig:systemGeometry}. We seek to image an object at a range $R_c$ relative to the observing camera. If the camera directly observes the object, the minimum object feature resolvable is equal to the instantaneous field of view (ifov), $\Delta R_c/F$, where $\Delta$ is the detector pitch and $F$ is the focal length~\cite{brady2009optical}. We assume that $\Delta\geq \lambda f/\#$, where $\lambda$ is the operating wavelength and $f/\#$ is the f-number, to ensure consistency with the diffraction limit. Alternatively, the camera may choose to observe light scattered between the object and the camera instead of the object itself. Observable scatter may arise from air or water borne particles or from secondary reflections off of intermediate surfaces. Recognizing that the scatter may occur over a three dimensional region, here we assume for simplicity that the scatter is isolated at a single plane. Volume scatter would require tomographic phase retrieval but would not otherwise affect our resolution argument. We also assume that the scattering process is incoherent, in which case a random phase is imposed on the scattered field. While this means that phase sensitive or holographic detection of the scattered field is unlikely to be useful, incoherent scattering is essential to ensuring that the scattered field is observable at the camera aperture (e.g. the scatterer radiates uniformly in all directions). 

We assume that the object is illuminated by a known coherent source. Representing the field on the scattering surface as $\psi(x)$, the camera images $|\psi|^2$, i.e. the radiant intensity of the scatter. To backproject the scatter data into an image of the object we need to recover the complex field $\psi(x)$. To achieve this objective, we use relative motion between the object and the scattering plane to obtain diverse measures of $|\psi|^2$ and then we apply iterative algorithms with prior information to recover the phase. Prior information may include, for example, the trajectory of the relative motion, the spatial support of the object or object surface characteristics.

Assuming that we are able to recover $\psi$ on the surface of the scatterer, the object is imaged by computational backpropagation. In practice, backpropagation is implemented as part of the phase recovery process by iterating between scatter space and object space with a support constraint on the object. In this process, the minimum resolvable object feature is $\lambda R_s/A_s$, where $R_s$ is the range between the object and the scattering plane and $A_s$ is the cross-section of the scatter pattern. Comparing with the direct view ifov, we find the net resolution has improved by the ratio 
\begin{equation}
\label{eq:alpha}
    \alpha=\frac{\Delta A_s R_c}{\lambda R_s F}.
\end{equation}

The camera ifov remains important in determining the field of view on the object. The maximum reconstructed field of view (FOV) is equal to the ratio of the wavelength to the sampling period on the scatter. Assuming that the camera is diffraction limited this yields ${\rm FOV}=\frac{A_c}{R_{sc}}$, where $R_{sc}$ is the range between the camera and the scatter and $A_c$ is the camera aperture. Various illumination, motion and multiplane sampling strategies may be imagined to increase this field of view. For example, camera motion or a camera array could be used to synthesize a larger aperture as discussed in~\cite{wang2022snapshot}.
Here, however, we limit our focus to a simple demonstration using a single imaging aperture. 

\section{System Design}
\label{sec:system-design}
\begin{figure*}
\centering
\subfigure{\includegraphics[width=.8\linewidth]{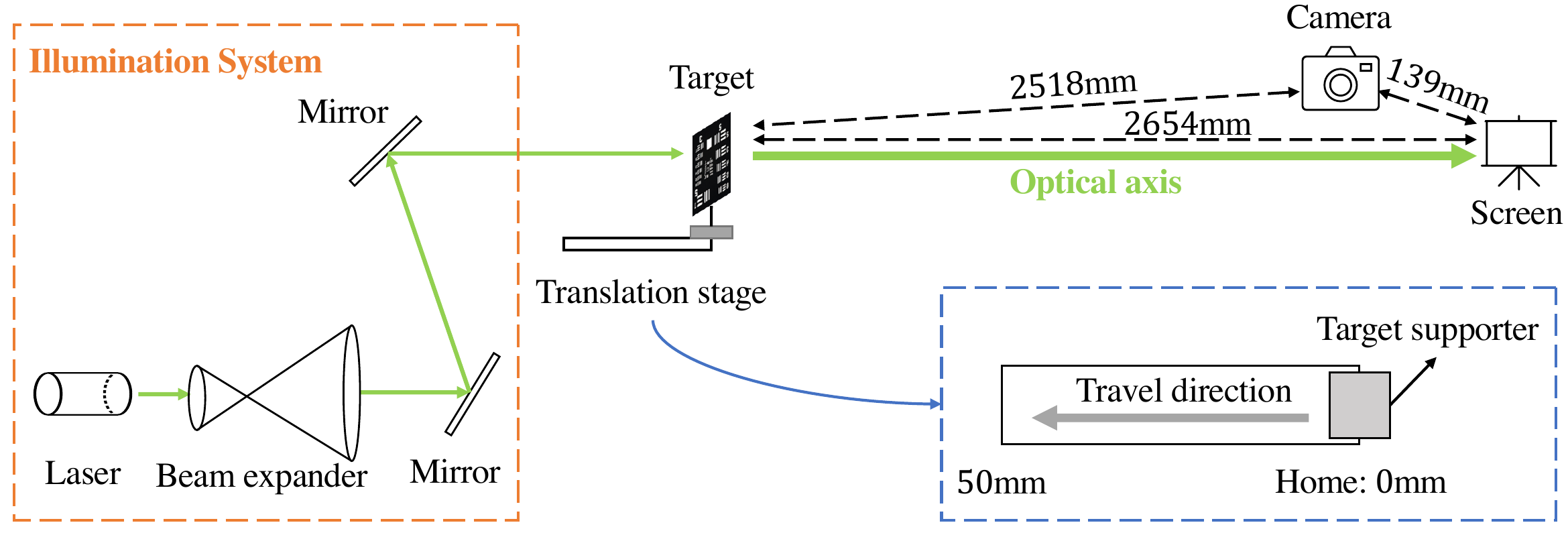}}
    \caption{\label{fig:drawing} 
    System layout. From left to right are: a collimated illumination system that produced a coherent planar wave, a transmissive planar target on the translation stage, a camera and a paper screen. Distances were measured when the target was at the home position (0 mm) on the stage.}
\end{figure*}

We used the system sketched in Fig.~\ref{fig:drawing} to demonstrate scatter ptychography. %The scatter was observed on a white paper screen. The distance between the screen and the object was 2654 mm and the scatter aperture was $\approx 38$ mm, corresponding to an estimated minimum feature resolution of $\frac{\lambda R_S}{A_s}\approx 71 $ microns at $\lambda = 0.532$ microns. The range between the camera and the target was 2500 mm. The camera focal length was 12 mm and the camera pixel pitch was 6.9 microns. With the camera used in this system the each pixel corresponds to a resolution of 1.4 mm on target. Thus, for this system the expected resolution improvement factor is $\alpha=38.7$. Our goal is simply to show that by observing and processing the scattered field we can resolve $\approx 40$ micron features, thus improving on the direct view camera resolution at this range by $\approx 36\times$
The illumination was a $\lambda = 532$ nm collimated continuous-wave laser beam. This plane wave illuminated a planar target mounted on a translation stage. The stage allows up to 50 mm of travel along optical axis. The target was positioned within the uniform region of the incident beam, perpendicular to the optical axis. We let $f(x,y)$ denote the modulated field right after the target. A paper screen was placed perpendicular to the optical axis, $R_s = 2654$ mm away from the home position of the target. We mounted a 12 mm F1.6 lens on a monochromatic camera, which was $R_{sc} = 139$ mm in front of the screen. The detector of the camera had $540\times720$ resolution and $\Delta = 6.9$ microns detector pitch, corresponding to  $\frac{\Delta R_{sc}}{F} \approx 80$ microns ground sample distance on the screen.

\begin{figure}[htbp]
\centering
\subfigure[Resolution chart]{
  \includegraphics[width=.23\linewidth]{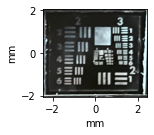}
}
\subfigure[Scatter image (0 mm)]{
  \includegraphics[width=.22\linewidth]{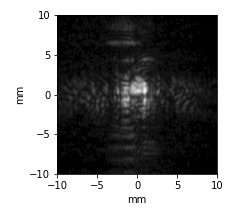}
}
% \subfigure[25mm]{ 
%   \includegraphics[width=.22\linewidth]{figs/scatter25mm.png}
% }
\subfigure[Scatter image (50 mm)]{
  \includegraphics[width=.22\linewidth]{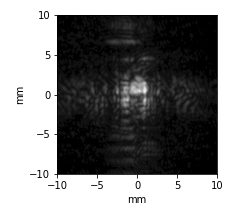}
 }
\subfigure[Difference map]{
  \includegraphics[width=.24\linewidth]{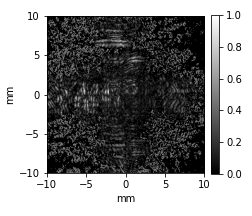}
}
% \vspace*{-3mm}
\caption{The resolution chart (a) and two of its scatter images when the target was at (b) 0 mm and (c) 50 mm on the stage. We cropped both scatter images and illustrated their square roots. (d) shows the normalized absolute differences between (b) and (c), where the whiter region indicates more derivation.}
\label{fig:usaf_and_scatters}
\end{figure}

We adjusted the exposure time to avoid saturation in recording the scatter image $|\psi|^2$ on the screen. 
%We recorded and averaged 500 images to increase the dynamic range of the signal. 
We selected the point where the optical axis and the screen intersected as the center of the camera image, then cropped the recorded image to the central up to $460\times 460$ pixel region. We used simple geometric analysis to project the camera image onto screen coordinates. The above procedure of data preprocessing was repeated for two different relative displacements of the target and scatter planes. Fig.~\ref{fig:usaf_and_scatters}(a) shows a 1951 USAF resolution chart, the dimension of which was 5 mm $\times$ 4 mm. Fig.~\ref{fig:usaf_and_scatters}(b) and (c) show the central portion of the scatter images at different object positions, while the cross-section $A_s$ of the full scatter images within the camera FOV was around 37 mm, corresponding to an estimated minimum feature resolution $\frac{\lambda R_s}{A_s}\approx 38$ microns at $\lambda = 0.532$ microns. Differences between two scatter images are illustrated in Fig.~\ref{fig:usaf_and_scatters}(d). %From measurements above we expect the minimum feature resolution to be $\frac{\lambda R_s}{A_s}\approx 38$ microns. 

The range between the camera and the home position of the target was 2518 mm. With the camera used in this system, each pixel {corresponded} to a resolution of 1.45 mm on target. Thus, for this system, the expected resolution improvement factor is $\alpha\approx38$. As discussed below, we experimentally resolved resolution chart features at $\approx 44$ micron, thus improving on the direct view camera resolution at this range by $\approx 32\times$.
\section{Image Estimation}
We used error-reduction based phase retrieval~\cite{bauschke2002phase} to image the object given a set of scatter images $|\psi_1|^2$ to $|\psi_{n_p}|^2$ and associated distances $z_1$ to $z_{n_p}$. To model this process, consider the two-dimensional complex field right after the target surface be represented as $f(x, y)$ and its Fourier transform as $\hat{f}(u, v)$, where the vector $(x, y)$ is a spatial coordinate and $(u, v)$ a spatial frequency. The propagated complex field $\psi(x, y)$ on the screen surface a distance $z$ to the target can be modeled by scalar diffraction as
\begin{equation}
\label{eq:etf}
    \psi(x, y) = \iint^{\infty}_{-\infty} \hat{f}(u, v)\exp{\left(j2\pi z\sqrt{\frac{1}{\lambda^2}-u^2-v^2} \right)} \times \exp{\left[j2\pi(ux + vy) \right]} \,du\,dv,
\end{equation}
\noindent where $\sqrt{u^2 + v^2} < 1/\lambda$ for propagating waves. Note that one can also calculate $f$ from $\psi$ and its Fourier transform $\hat{\psi}$ by backpropagation ($z < 0$). Hence estimating the target $f$ is equivalent to estimating the phase of $\psi$.  Phase retrieval can be achieved by many different sampling and processing strategies~\cite{shechtman2015phase}, here we apply an error-reduction algorithm in combination with numerical diffraction between the object and multiple scatter planes.

Assuming in Eq.~\ref{eq:etf} $f(x, y)$ is sampled on a Cartesian grid with period $\delta$ and is zero-padded to have $N$ samples in the x and y dimensions, the diffracted field may be calculated by discrete Fourier transforms with the angular spectrum transfer function~\cite{Zhang:20}. The sampling periods for the DFT are $\Delta x = \Delta y = \delta$, and $\Delta u = \Delta v = 1/(N\delta)$. Let $[m, n]$ be a discrete spatial coordinate and $[p, q]$ be a discrete frequency coordinate, all running from $-N/2$ to $N/2$. The diffracted field is:
\begin{equation}
\psi[m, n] = \text{DFT}^{-1}\left\{\text{DFT} \left\{f[m, n]\right\} \exp{\left[j 2\pi z\sqrt{\frac{1}{\lambda^2} - \frac{p^2 + q^2}{(N\delta)^2}}\right]} \right\}.
\end{equation}
%where $\text{FFT}$ represents Fast Fourier Transformation and $\text{FFT}^{-1}$ represents its inverse. 

Assuming that the field is most tightly focused at the object, the extent of the field expands on propagation. With a constant space-bandwidth product, this means that the transverse spatial frequencies of the field decrease on propagation (e.g. the field blurs). When the field propagates forward ($z > 0$), the frequency decreases by the rate proportional to $1/z$. Once the Nyquist rate of the field drops significantly below the sampling rate, the field can be downsampled without incurring aliasing. Similarly, the field during backward propagation ($z < 0$) can be upsampled when appropriate. In light of this, we use a multistage angular spectrum method (MASM) with bicubic down/up-sampling of the field upon significant decreases/increases in the spatial bandwidth. Here a "stage" refers to a single resampling behavior and "multi" indicates resampling can happen more than once. Let $f$ be sampled by the aforementioned grid and $\beta$ ($0<\beta<1$) be the ratio of downsampling, we have $\psi[m', n'] = \texttt{MASM}\{f[m, n], z\}$ and $f[m, n] = \texttt{MASM}^{-1}\{\psi[m', n'], z\}$, where $m'$ and $n'$ runs from $-\beta N/2$ to $\beta N/2$ and $\Delta m' = \Delta n' = \delta/\beta$.

\texttt{MASM} and its inverse lift the constraint that $f$ and $\psi$ share the same sampling rate and enable propagation within our computational budget. In our experiments, $\delta$ was 10 microns and $\beta$ was 0.25. We applied MASM analysis in the phase retrieval algorithm shown in Alg.~\ref{alg}.%, which can be interpreted as a multiplane Gerchberg-Saxton algorithm. 

%Based on the known structure of the objects under test, the algorithm applied prior constraints on $f[m, n]$. First, we applied a support constraint based on the knowledge of the location of the object. Support was limited to a rectangular region consisting of $a$ pixels in $m$ direction and $b$ pixels in $n$ direction. %Due to the ambiguity of the phase retrieval algorithm~\cite{goodman2005introduction}, the true field $f$, $f$ with a constant phase shift and $f^\ast$ reflected about the origin are all acceptable solutions. We removed this ambiguity by requiring $f$ to be real and non-negative based on the knowledge the objects under test are amplitude masks.

The algorithm started with random initialization of $f$, and improved the estimate recursively. Here we assigned a random phase $\varphi$ to one of the amplitude measurements, $|\psi_1|$ for example, and backpropagated it to derive our first estimate of $f$. For one plane $c_k$ which was selected randomly, the estimate of $f$ was made to conform to {a support constraint based on the knowledge of the location of the object. Support was limited to a rectangular region consisting of $a$ pixels in $m$ direction and $b$ pixels in $n$ direction. The restrained field was propagated a distance $z_{c_k}$ forward .} %with only amplitude values within its region of support. 
The propagated field $\psi_{c_k}$ kept the phase but replaced its amplitude with the known measurement. Then the new field propagated inversely, yielding a new estimate of $f$. Once all the planes were visited, one iteration was finished and the estimate of $f$ would be the start point of the next, until the maximum number of iterations was reached.

\begin{algorithm}[htbp]
\small
\caption{scatter ptychography algorithm}\label{alg}
\begin{algorithmic}
\Require $n_p$\Comment{number of planes}
\State $\Psi = \{|\psi_1[m', n']|^2, |\psi_2[m', n']|^2, \dots, |\psi_{n_p}[m', n']|^2 \}$ \Comment{scatter images}
\State $Z = \{z_1, z_2, \dots, z_{n_p} \}$ \Comment{propagation distances}
% \State $C[m, n]$ \Comment{light field at the target}
\State $a, b$ \Comment{target size in pixels along m and n axes}
\State $n_i$ \Comment{number of iterations}
\\
\State $\varphi \sim \mathcal{N}(\mathbf{0}, \mathbf{1}\cdot \mathbf{1}^T)$\Comment{sample from the standard normal distribution}
\State $f[m, n] \gets \texttt{MASM}^{-1}\left[|\psi_1[m', n']|e^{j\varphi}, z_{1}\right]$\Comment{initial estimation of the target}%/C[m, n]$ 
\State $i \gets 0$
\While{$i < n_i$}
    \State ${c_1, c_2, \dots, c_{n_p}} \gets \texttt{shuffle} \{1, 2, \dots, n_p\}$
    \For{\texttt{each} $c_k$}
        \State $f[m, n] \gets \texttt{rect}(m/a) \texttt{rect}(n/b) f[m, n]$ \Comment{support constraint}
        \State $\psi_{c_k}[m', n'] \gets \texttt{MASM}\left[f[m, n], z_{c_k}\right]$ \Comment{forward multistage propagation}
        \State $\psi_{c_k}[m', n'] \gets |\psi_{c_k}[m', n']|e^{j\phi\left[\psi_{c_k}[m', n']\right]}$ \Comment{update phase estimation} 
        \State $f[m, n] \gets \texttt{MASM}^{-1}\left[\psi_{c_k}[m', n'], z_{c_k}\right]$ \Comment{backward multistage propagation}
    \EndFor
    \State $i \gets i+1$
\EndWhile

\noindent\Return $f[m, n]$
\end{algorithmic}
\end{algorithm}
\noindent where $|\cdot|$ and $\phi(\cdot)$ take the amplitude and phase of a complex matrix elementwisely, $\mathbf{1}$ is a column vector $\in \mathbb{R}^{\beta N}$ of all 1 entries, and \texttt{rect}(x) is the rectangle function:
\begin{equation}
\texttt{rect}(x)=
\begin{cases}
  1, & \ |x| < \frac{1}{2} \\
  \frac{1}{2}, & \ |x| = \frac{1}{2} \\
  0, & \text{otherwise}
\end{cases}.
\end{equation}
%, equal to 1 if $|x| < 0.5$, 0.5 if $|x| = 0.5$, and 0 otherwise. 
%Compared to the iterative multiple plane parallel algorithm [cite], our algorithm updates $f$ estimate once it visits a plane instead of averaging estimates from all planes. In practice we found our algorithm yields sharper results.
\section{Experimental Results}
\label{sec:results}
The aforementioned resolution chart was used as the target to analyze the resolution of our proposed system.
As a check on our analysis, we also simulated our experimental system including target modeling, propagation of the target field, and projection of scatter images. We set the number of planes $n_p = 2$, corresponding to 0 mm and 50 mm on the stage. For each plane, the field of the target was propagated the associated distance to a synthetic screen, where only the intensity was kept. The intensity image was projected onto the camera sensor according to the geometry.%a model including the focal length and relative orientation. %Preprocessing of synthetic data only involved reprojection, as in simulation images represented by 64-bit floating numbers are of high dynamic range, thus no averaging was necessary. 

\begin{figure}[htbp]
\centering
\subfigure[Synthetic target]{
  \includegraphics[width=.28\linewidth]{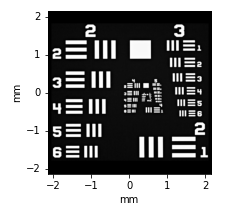}
}
% \subfigure[Scatter image at 25mm]{
%   \includegraphics[width=.2\linewidth]{experimentFigures/25mm_sim.png}
% }
\subfigure[Synthetic reconstruction]{
{\includegraphics[width=.33\linewidth]{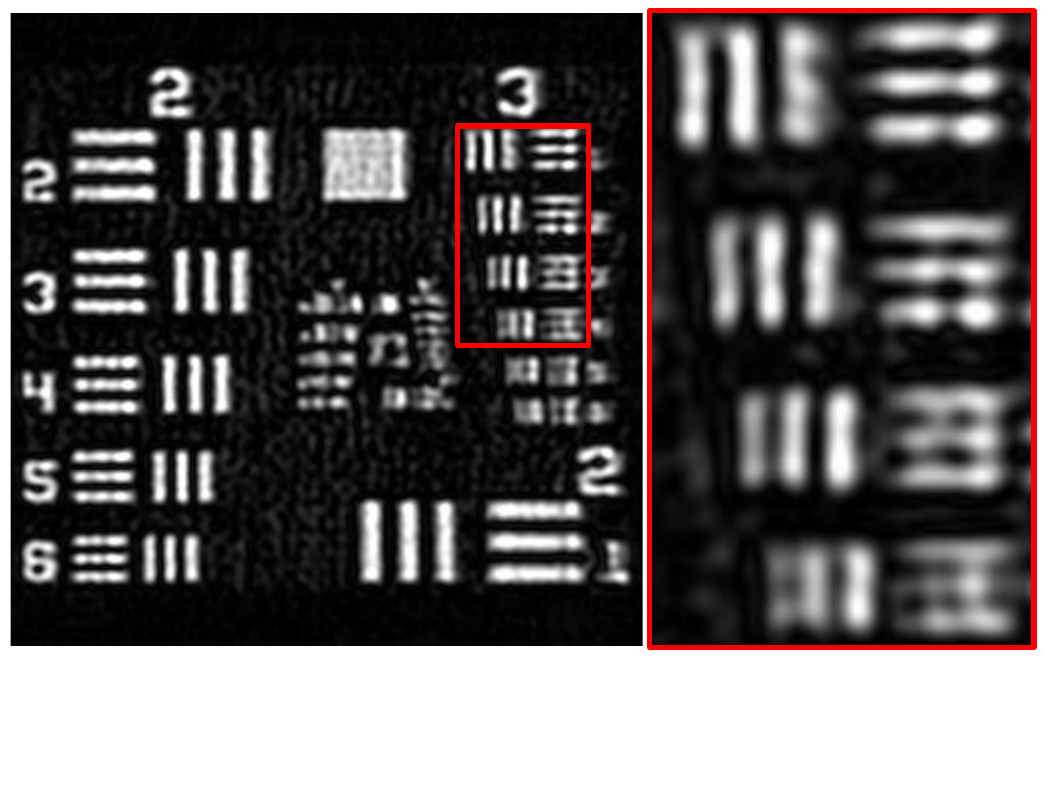}}
}
\subfigure[Experimental reconstruction]{
{\includegraphics[width=.33\linewidth]{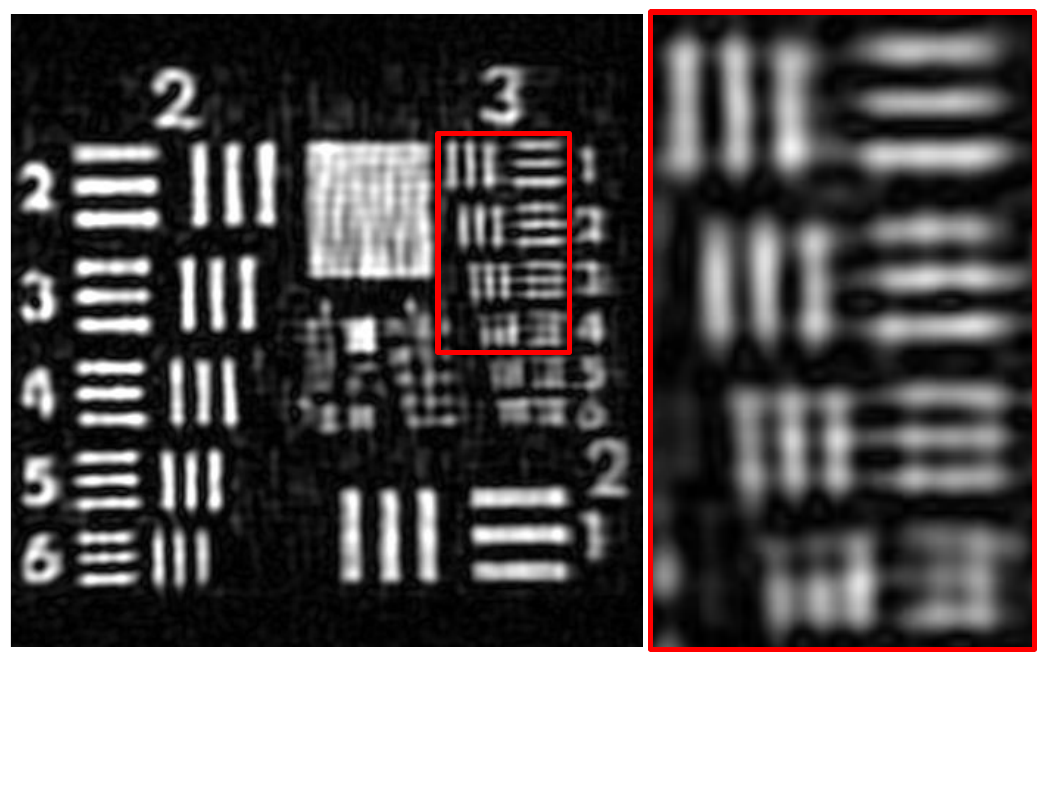}}
}
% \vspace*{-3mm}
\caption{The synthetic target and reconstructions from synthetic and experimental scatter images. Images were zero-padded to $430\times430$ resolution, in which each pixel represented 10 microns physical spacing. In (b) and (c), elements 1 to 4 of group 3 in the red blocks were magnified and displayed by the side.}
\label{fig:sim}
\end{figure}

Then synthetic camera images were preprocessed as stated in section~\ref{sec:system-design} and fed into the phase retrieval algorithm. {
Due to the ambiguity of the phase retrieval algorithm~\cite{goodman2005introduction}, the true field $f$, $f$ with a constant phase shift and $f^\ast$ reflected about the origin are all acceptable solutions. We removed this ambiguity by requiring $f$ to be real and non-negative based on the knowledge the objects under test are amplitude masks. Hence we enforced this constraint in the following experiments on top of the support constraint by setting $f[m, n] \gets \texttt{rect}(m/a) \texttt{rect}(n/b) |f[m, n]|$}.
%We set the support region $a = 360,b = 412$ for both, corresponding to 3.6mm $\times$ 4.12mm real dimensions. 
The target used in the simulation is shown in Fig.~\ref{fig:sim}(a). The algorithm converged within $n_i = 200$ iterations.  Fig.~\ref{fig:sim}(b) shows a simulated reconstruction. Blur relative to the true target arises from the limited scatter aperture, we consider additional blur due to photon noise in section~\ref{sec:discussion}. Applying the same processing (except for $n_i = 500$) to actual scatter data yields the result shown in Fig.~\ref{fig:sim}(c). Reconstructions from synthetic data and real data are consistent, resolving up to the 4th element of group 3 (11.31 lp/mm). This is equivalent to $44.2$ mm ground sample distance.

\begin{figure}[htbp]
\centering
\subfigure[1 plane (0 mm)]{
  \includegraphics[width=.21\linewidth]{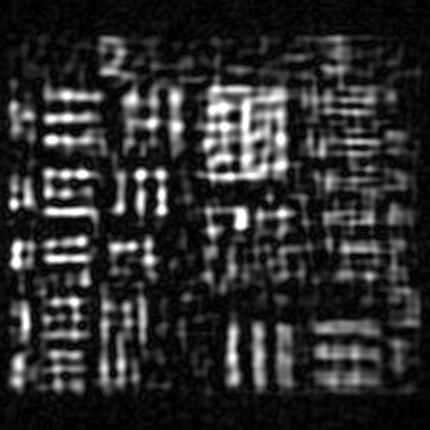}
}\hspace{0.05\linewidth}
\subfigure[1 plane (50 mm)]{
  \includegraphics[width=.21\linewidth]{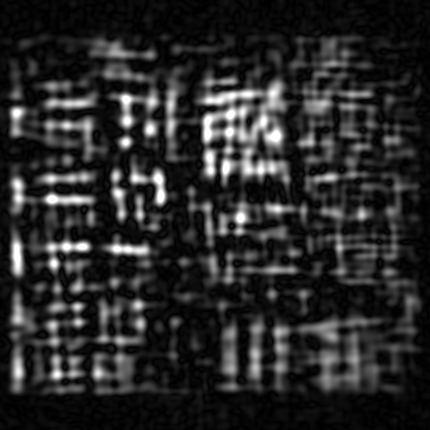}
}\hspace{0.05\linewidth}
\subfigure[2 planes (0, 50 mm)]{
\includegraphics[width=.21\linewidth]{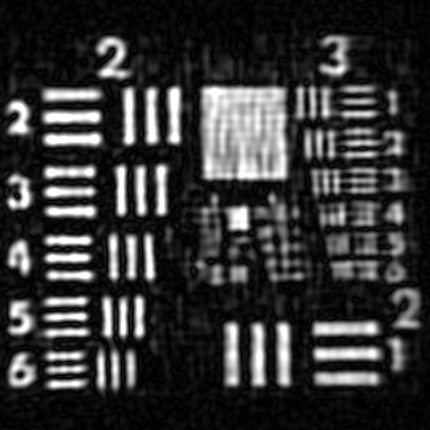}
}
% \vspace*{-3mm}
\caption{Reconstructions with 1 plane and 2 planes. Numbers in the parenthesis indicate the associated target positions on the stage.}% {\color{red} crop a bit and magnify to make the reconstructions clearer. } }
\label{fig:numofplanes}
\end{figure}

One may wonder if a comparable result can be achieved using a single plane. %The following experiment was to investigate the optimal number of planes.
To investigate this problem, we carried out experiments on $n_p$ while keeping other parameters of the algorithm the same. %and the number of iterations $n_i$. For settings with more than one plane, the positions of the target spread out evenly in the 50 mm range. 
We chose $n_i = 500$ to ensure convergence. From Fig.~\ref{fig:numofplanes}, we observe that the result drastically deteriorated with only one plane. We also tested reconstructions with more than two scatter planes but for this geometry, target and constraint we did not observe significant improvement.
Based on these observations, $n_p$ was set to 2 in the following experiments with the target at positions 0 and 50 mm.

%It seems reasonable that a single plane cannot produce the best result, but counterintuitive at first sight that more data does not necessarily produce better results. There are two reasons: 1) more data may contribute more to redundancy and less to the information; 2) more noises and errors are likely to be introduced by technical operations.  

\begin{figure}[htbp]
\centering
\subfigure[Target "A"]{
  \includegraphics[width=.32\linewidth]{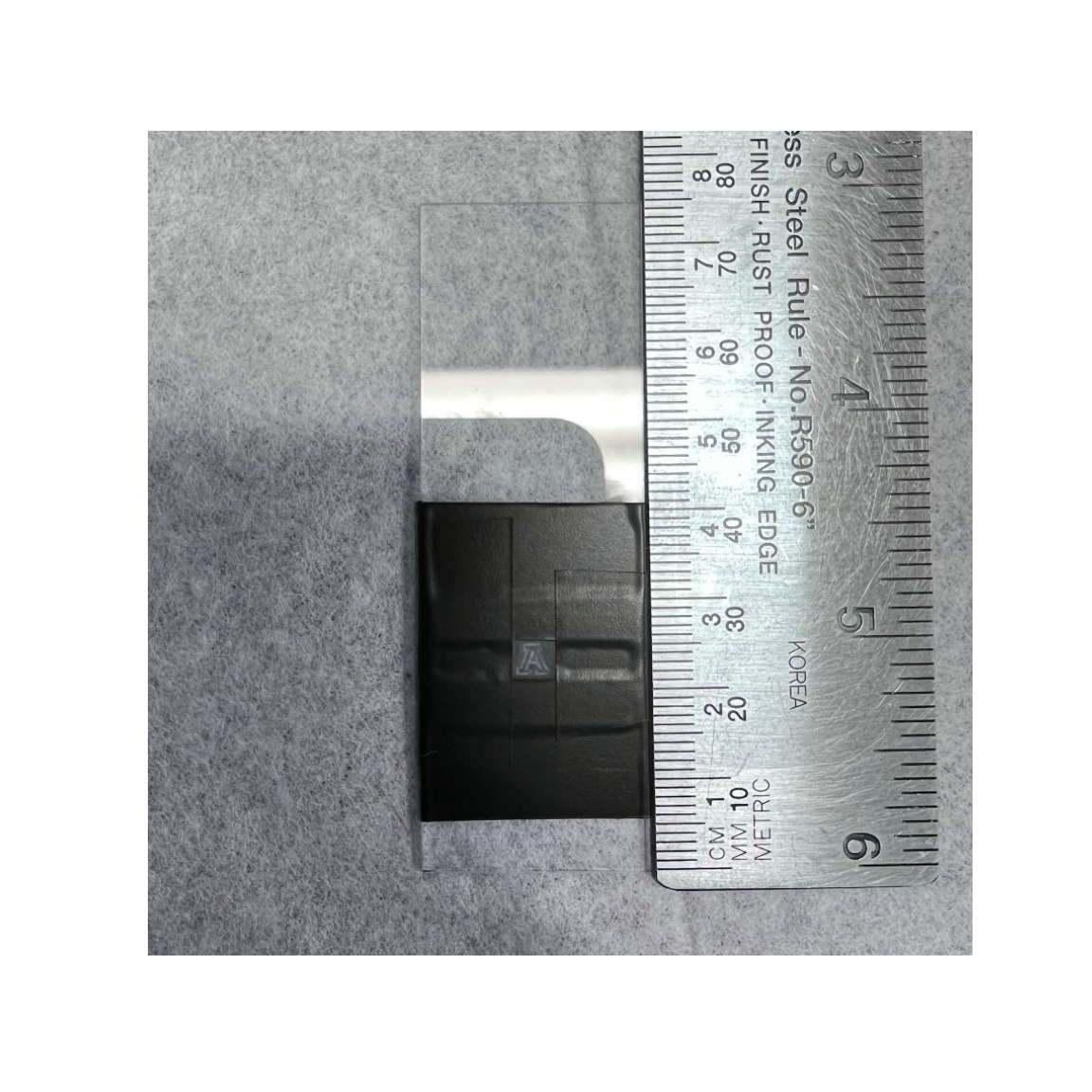}
}
\subfigure[Scatter image (0 mm)]{
  \includegraphics[width=.32\linewidth]{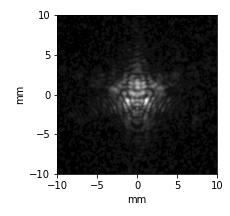}
}
% \vspace*{-3mm}
\caption{Target "A" and one of its scatter images, captured when the target was at the home position. We cropped the scatter image and illustrated its square root.}
\label{fig:sample}
\end{figure}

To analyze system performance for diverse objects, we printed additional targets on plastic transparency film. Three targets with text "A", "OSC" and "UoA" were $3\sim 5$ mm in height and $4\sim 7$ mm in width and printed clear against the black background. Each target was cropped and mounted on a  microscope slide. Fig.~\ref{fig:sample} shows one of the targets and its scatter image. 

\begin{figure*}[htbp]
\centering
{
\begin{minipage}[c]{1.0\linewidth}
\centering
\setlength\tabcolsep{.5pt}
\begin{tabular}{ccccc} 
\toprule
& Resolution chart & "A" & "OSC" & "UoA" \\
\midrule
Target & 
\subfigure{
\includegraphics[height=0.15\textwidth]{experimentFigures/usaf_cropped_axison.png}}		&
\subfigure{
{\includegraphics[height=0.15\textwidth]{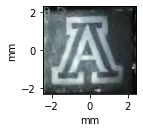}}}		&\subfigure{
{\includegraphics[height=0.15\textwidth]{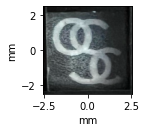}}}		&\subfigure{
{\includegraphics[height=0.15\textwidth]{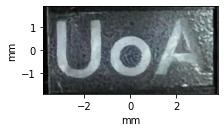}}}		\\
\midrule
Result & 
\subfigure{
{\includegraphics[height=0.12\textwidth]{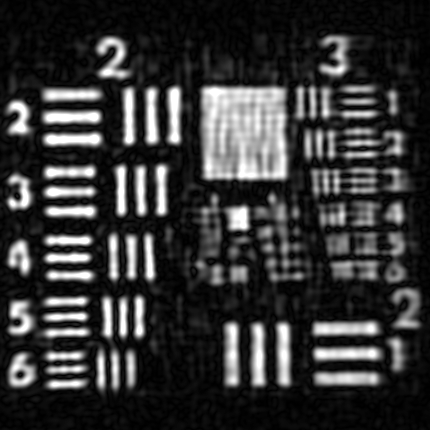}}}		&
\subfigure{
{\includegraphics[height=0.12\textwidth]{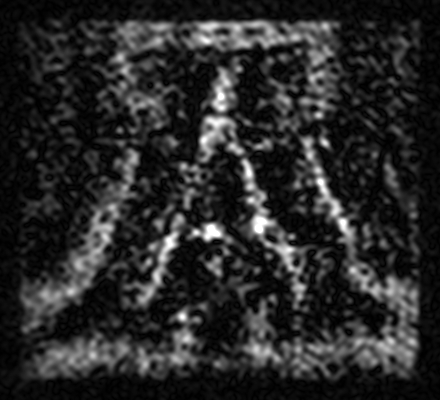}}}		&\subfigure{
{\includegraphics[height=0.12\textwidth]{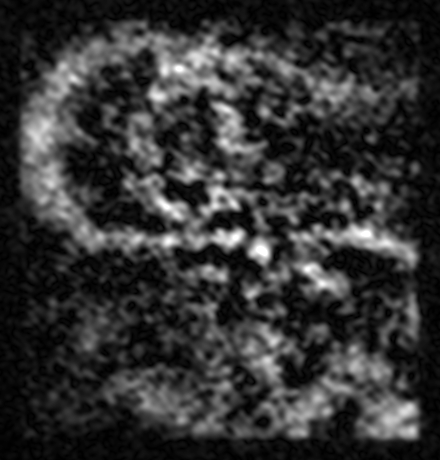}}}		&\subfigure{
{\includegraphics[height=0.12\textwidth]{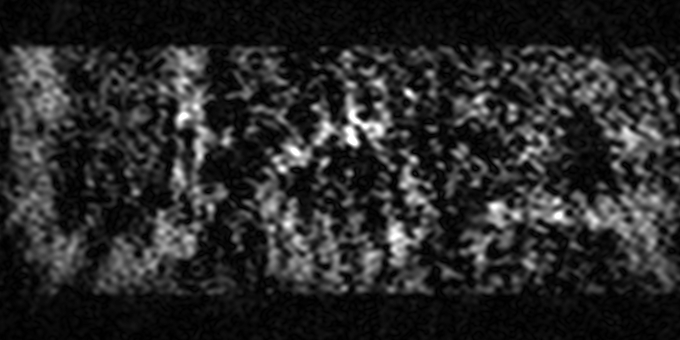}}}		\\
\midrule
Direct view & 
\subfigure{
\scalebox{-1}[1]{\includegraphics[height=0.12\textwidth]{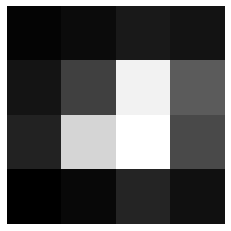}}}		&
\subfigure{
\scalebox{-1}[1]{\includegraphics[height=0.12\textwidth]{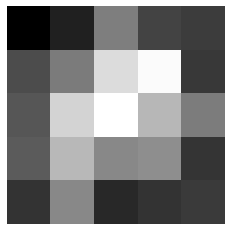}}}		&\subfigure{
\scalebox{-1}[1]{\includegraphics[height=0.12\textwidth]{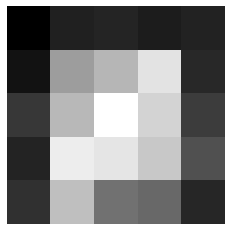}}}		&\subfigure{
\scalebox{-1}[1]{\includegraphics[height=0.12\textwidth]{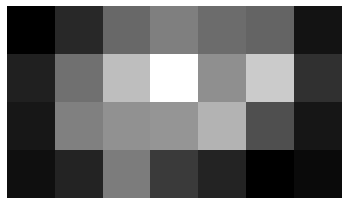}}} \\
\bottomrule
\end{tabular}
\end{minipage}
\caption{\label{fig:comptable}Comparison between direct views and our reconstruction results of 4 targets. Images were cropped to highlight the regions of interest, and rescaled for better alignment in the table. The actual size of each target is labeled, which also applies to its reconstruction and direct view. The pixel pitch of reconstructions was 10 microns, while of direct views was 1.42mm.}
}
\end{figure*}

We reconstructed each target from two of its associated scatter images. The estimate of each target converged in 200 iterations. Due to inhomogeneity in the plastic films, the constraint that $f$ is real was not strictly valid, thus outputs are not as sharp as for the chrome on the glass resolution target. Nevertheless, comparison in Fig.~\ref{fig:comptable} between direct views and reconstruction results from our phase retrieval algorithm demonstrates significant improvements in resolution. %Direct views are captured by twisting the camera to look at the targets at the origin directly, thus are horizontally flipped compared to the reconstructions. For ease of comparison, these images are flipped back in the figure. 
From calibration we measured that each pixel of the direct view corresponded to 1.42 mm on the target at the home position, indicating the net resolution in practice was improved by the ratio $\alpha = \frac{1.42\ \text{mm}}{44.2\ \text{um}} \approx 32$.

\section{Discussion}
\label{sec:discussion}
As we have seen, scatter ptychography can be used to improve imager resolution by more than an order of magnitude over direct view performance. Of course, the critical question is how large can the improvement factor, $\alpha$, as defined in Eqn.~\ref{eq:alpha}, be? To answer this question, we consider the scatter signal generated by object features at the resolution limit. An object feature of cross-section $\delta$ reflects radiant power $P\delta^2$, where $P$ is the power density illuminating the object. The fraction of this power collected by the camera is $\sigma\frac{A_c^2}{4\pi R_{sc}^2}$, where $\sigma$ is the fraction of the radiant power that is scattered and $R_{sc}$ is the range between the scattering surface and the camera. As discussed above, feature size is related to $A_s$ by $\delta\approx \lambda R_s/A_s$. Assuming that a detectable feature must deliver $N_p$ photons to the camera, one finds  $A_s\leq \frac{\lambda R_s A_c}{2R_{sc}}\sqrt{\frac{\sigma TP}{\pi N_p}}$, where T is the exposure time.

To get an idea of the limits of this relationship, one might assume that the illuminating power density is limited by diffraction from the observing aperture, i.e. $P=\frac{LA_c^2}{\lambda^2 R_C^2}$, where $L$ is the power of the illuminating source. In this case
$A_s\leq \frac{ R_s A_c^2 }{2R_{sc}R_c}\sqrt{\frac{\sigma TL}{\pi N_p}}$, where $T$ is the exposure time, 
and
\begin{equation}
    \alpha \leq \frac{ \Delta}{\lambda f/\# }\frac{ A_c}{2 R_{sc} }\sqrt{\frac{\sigma TL}{\pi N_p}},
\label{eq:alphaV}
\end{equation}
$\alpha$ separates into three interesting factors, (1) the pixel pitch relative to the diffraction limit, $\Delta/(\lambda f/\#)$, (2) the field of view of the camera on the scatter, $A_c/R_{sc}$, and (3) the inverse root of the quantum efficiency for collection of illuminating photons. The first factor one may expect ideally to be $\approx 1$.
Since one generally expects that $\frac{A_c}{R_{sc}} \ll 1$, $\alpha >1 $ requires $\sqrt{\frac{\sigma TL}{\pi N_p}}\gg 1$. 

\begin{figure}[htbp]
\centering
\subfigure[0.01 photons per pixel]{
{\includegraphics[width=.25\linewidth]{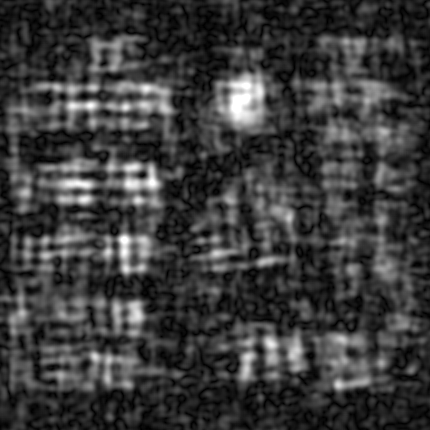}}
}\hspace{0.11\linewidth}
\subfigure[0.02 photons per pixel]{
{\includegraphics[width=.25\linewidth]{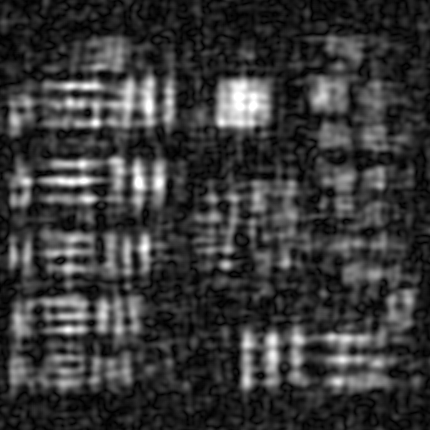}}
}\hspace{0.11\linewidth}
% \subfigure[0.03]{
% {\includegraphics[width=.17\linewidth]{figs/15photons.png}}
% }
% \subfigure[0.05]{
% {\includegraphics[width=.17\linewidth]{figs/25photons.png}}
% }
\subfigure[0.1 photons per pixel]{
{\includegraphics[width=.25\linewidth]{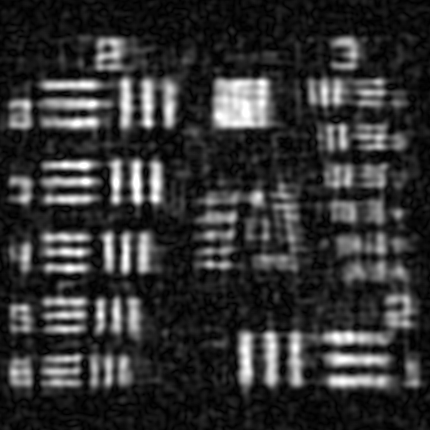}}
}\hspace{0.11\linewidth}
\subfigure[1 photons per pixel]{
{\includegraphics[width=.25\linewidth]{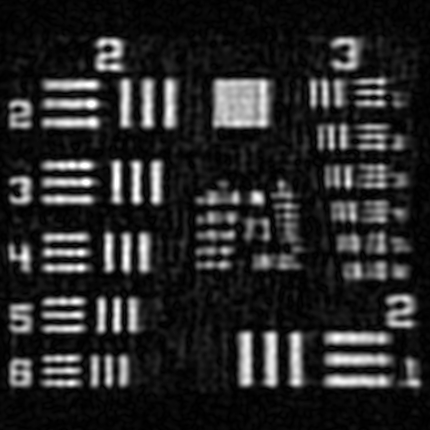}}
}
% \vspace*{-3mm}
\caption{Simulated reconstruction vs. exposure. Captions represent the average photons per pixel of the scatter images.}
\label{fig:expo-vs-res}
\end{figure}

We explored the relationship between the reconstruction resolution and the exposure in simulation. Fig.~\ref{fig:expo-vs-res} shows results using same simulation strategy discussed in section~\ref{sec:results} with the addition of Poisson noise for various exposure levels. The exposure level is listed in photons per pixel in the diffracted field, but intensity is not uniformly distributed in this field so expected flux at important features will greatly exceed $1$ photon. 
%{\color{blue}We applied same simulation strategy discussed in section~\ref{sec:results} except for averaging (i.e., we only used one scatter image per target position), ensuring the photon noise dominated the resolution.} 
At high flux levels $\alpha$ may be limited by the observed aperture limited, by as illustrated in~\ref{fig:expo-vs-res}(a) and (b) as flux drops the effective aperture will drop below the value defined by the camera field of regard. Eqn.~\ref{eq:alphaV} suggests that resolution should fall in proportion to the square root of the flux. Fig.~\ref{fig:expo-vs-res}(a) and (c) are roughly consistent with this prediction with with reduction from ~10 line pairs per millimeter (lp/mm) in (c) to ~4 lp/mm (a) for a $10\times$ reduction in flux. 

The fact that the illumination flux, $TL$, must greatly exceed the minimum detectable flux, $N_p$, in order for scatter ptychography to achieve an advantage is not surprising. The factor $\sigma\frac{A_c^2}{4\pi R_{sc}^2}$ reflects the loss in quantum efficiency of scatter imaging relative to a coherent aperture at the same location, this factor might easily be $<10^{-7}$. However, using coherent illumination it is not unreasonable to illuminate targets with flux that will overcome this loss. Since $TL$ may exceed $10^{20}$ photons and $N_p$ might be as little as $10^5$, it is not unreasonable to imagine $\sqrt{\frac{\sigma TL}{\pi N_p}}\approx 10^7$, which leads to diverse situations with $\alpha \gg 1$. For example, if the laser power is 1 KW and the observation time is 0.1 seconds, $TL\approx 10^{21}$. With this flux at illuminating a target at range 10 Km with the observing aperture of 10 cm and a scatter efficiency of $\sigma =0.1$, $\alpha$ might exceed $1000$.

\bibliographystyle{unsrt}  
\bibliography{references}  %%% Uncomment this line and comment out the ``thebibliography'' section below to use the external .bib file (using bibtex) .





% \end{thebibliography}

\end{document}